\documentclass[aps,prl,twocolumn,superscriptaddress,amsfonts,showpacs]{revtex4-1}
\usepackage{graphicx,color}
\usepackage{bm}
\usepackage[latin1]{inputenc}
\usepackage{hyperref}

\begin{document}

\title{High On/Off Ratios in Bilayer Graphene Field Effect Transistors Realized by Surface Dopants}

\author{B. N. Szafranek}
\email{szafranek@amo.de} \homepage{http://www.amo.de}
\author{D. Schall}
\author{M. Otto}
\affiliation{Advanced Microelectronic Center Aachen (AMICA), AMO GmbH, Otto-Blumenthal-Strasse 25, 52074 Aachen, Germany}
\author{D. Neumaier}
\affiliation{Institute of Semiconductor Electronics, RWTH Aachen University, Sommerfeldstr. 24, 52074 Aachen, Germany}
\affiliation{Advanced Microelectronic Center Aachen (AMICA), AMO GmbH, Otto-Blumenthal-Strasse 25, 52074 Aachen, Germany}
\author{H. Kurz}
\affiliation{Advanced Microelectronic Center Aachen (AMICA), AMO GmbH, Otto-Blumenthal-Strasse 25, 52074 Aachen, Germany}
\affiliation{Institute of Semiconductor Electronics, RWTH Aachen University, Sommerfeldstr. 24, 52074 Aachen, Germany}

\date{\today}

\begin{abstract}
The unique property of bilayer graphene to show a band gap tunable by external electrical fields enables a variety of different device concepts with novel functionalities for electronic, optoelectronic and sensor applications. So far the operation of bilayer graphene based field effect transistors requires two individual gates to vary the channel's conductance and to create a band gap. In this paper we report on a method to increase the on/off ratio in single gated bilayer graphene field effect transistors by adsorbate doping. The adsorbate dopants on the upper side of the graphene establish a displacement field perpendicular to the graphene surface breaking the inversion symmetry of the two graphene layers. Low temperature measurements indicate, that the increased on/off ratio is caused by the opening of a mobility gap. Beside field effect transistors the presented approach can also be employed for other bilayer graphene based devices like photodetectors for THz to infrared radiation, chemical sensors and in more sophisticated structures such as antidot- or superlattices where an artificial potential landscape has to be created. 
\end{abstract}

\pacs{}%
\keywords{}

\maketitle
%\section{Introduction}
The emergence of the 2-dimensional material graphene \cite{Geim,Heer,Geim2,Kim} has aimed a lot of research in the electron device community towards the realization of graphene based field effect transistors (FET). Creating a band gap to increase the on/off ratio of graphene FETs is one of the major challenges for using such devices in real applications \cite{Schwierz}. Therefore several concepts have been proposed, whereof lateral confinement \cite{Nakada,Wakabayashi} and the application of a perpendicular electric field to bilayer graphene \cite{McCann,Castro,McCann2} are probably the most promising ones. While there is currently no method to fabricate sufficiently narrow graphene nanoribbons by a top-down approach reliably, the situation is different for bilayer graphene with an electric field perpendicular to the basal plane. A band gap reaching values up to 250 meV has been observed by optical methods in micron-scale bilayer graphene devices \cite{Mak,Zhang}. However, increasing the on/off ratio in bilayer graphene FETs by a perpendicular electric field is more challenging as potential fluctuations and disorder enhance conductance in the off-state \cite{Oostinga,Zou,Taychatanapat} and thus reduce the on/off ratio. Recently quite high on/off ratios close to 100 at room temperature were achieved \cite{Xia,Szafranek}. In both experiments a top and a bottom gate were necessary to create a band gap and to vary the charge carrier concentrations simultaneously. From a technological point of view the control of a transistor's conductance by only one gate electrode is desired and hence one gate electrode has to be replaced. The creation of a band gap in bilayer graphene by adsorbate doping has already been observed by angle resolved photoemission spectroscopy \cite{Ohta} and by magnetotransport measurements at low temperatures \cite{Castro}. In the present work we show that this approach can be used to increase the on/off ratio in single-gated bilayer graphene FETs. Beside FETs the presented method has also large potentials for optoelectronic applications such as THz-detectors \cite{Ryzhii} or the creation of more sophisticated potential landscapes necessary in antidot lattices and superlattices \cite{Shen,Eroms}. In addition a clear understanding of adsorbate induced band gap modifications provides the scientific base for further development in the area of ultra sensitive sensors.

% Sample fabrication and characterization
In this work we fabricated back gated bilayer graphene field effect transistors, where the top side was freely accessible for adsorbate doping. We note that to observe the reported effects a proper substrate pretreatment and sample preparation is crucial. We used highly p-doped Si wafers covered with 90~nm thermally grown SiO$_2$ as a substrate. Prior to the graphene deposition the substrate was coated with hexamethyldisilazane (HMDS) in a chemical vapor deposition process. The HMDS coating makes the SiO$_2$ substrate highly hydrophobic and reduces the hysteresis and intrinsic doping concentration of graphene based FETs \cite{Lafkioti}. Subsequently, the graphene was exfoliated with an adhesive tape from a natural graphite crystal and deposited on the substrate. Bilayer graphene flakes were identified using optical microscopy and contrast determination of the graphene relative to the substrate \cite{Blake}. For some flakes Raman spectroscopy was used to verify their bilayer nature \cite{Ferrari}. After graphene deposition the contact electrodes were fabricated by optical lithography, sputter deposition of 40~nm nickel and a subsequent lift-off process. A schematic of the device stack is depicted in the inset of Fig.~1. 
We applied two different approaches to dope the upper graphene plane, one for p-type and one for n-type doping. The exposure of the fabricated graphene FETs to ambient atmosphere leads to a p-doping of the graphene. The doping is attributed to the absorption of water \cite{Schedin} or oxygen \cite{Ryu} molecules which are weakly bonded to the graphene surface. Although this method is not suited for device fabrication because of its experimental character, it is rather simple to perform and allows a stepwise removing of the dopants by annealing and hence a quasi continuous tuning of the doping level. Our second approach is based on thermal evaporation of a 1 nm thin aluminum layer, which oxidizes immediately when exposed to ambient atmosphere. Aluminum causes a quite strong n-type doping in graphene \cite{Giovanetti}. In both cases the doping effect on the graphene is caused by a charge transfer between the adsorbates and the graphene.

\begin{figure}[!t]
\includegraphics[width=\linewidth]{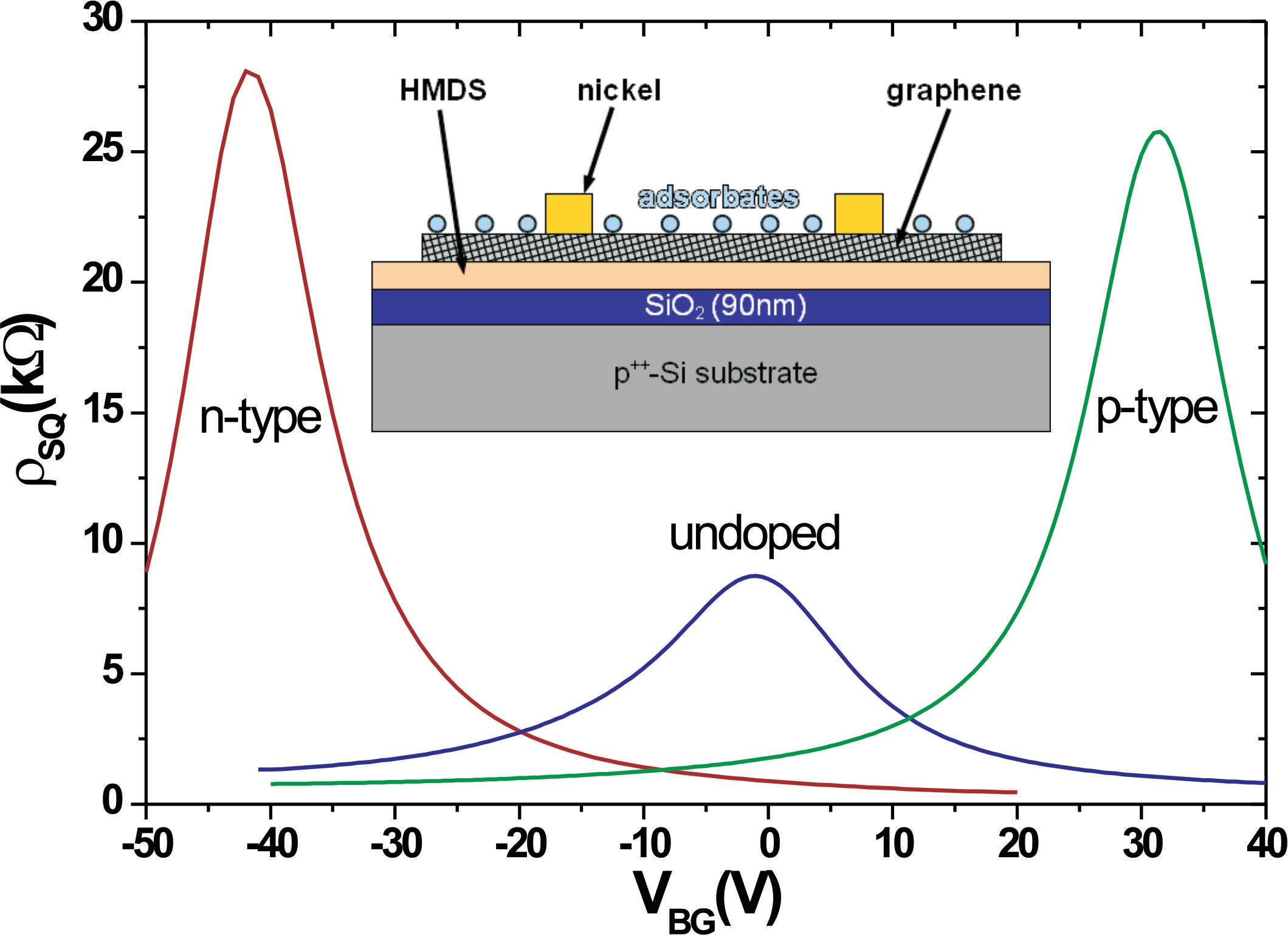}
\caption{Sheet resistance of a n-type, p-type and an undoped bilayer graphene FET as a function of the applied back gate voltage measured at room temperature in nitrogen atmosphere. The inset shows a schematic of a graphene FET. The doping by adsorbates is indicated.}
\end{figure}

The samples were measured in a needle probe station in nitrogen atmosphere (room temperature and above) and in vacuum (below room temperature). For electrical characterization a HP 4156 semiconductor parameter analyzer was used. The samples were characterized in a quasi four probe configuration with invasive contact electrodes at a constant source drain current of 10 $\mu$A. The sheet resistance $\rho_{SQ}$ as a function of the applied back gate voltage $V_{BG}$ of two freshly fabricated bilayer graphene devices with n-type (aluminum) and p-type (atmospheric) doping are depicted in Fig.~1 together with the characteristics of an undoped device, where atmospheric dopants were removed by an in-situ annealing at 200°C. The transfer characteristic of the undoped device is quite typical for a bilayer graphene FET without a band gap, having a sheet resistance $\rho_{SQ}$ of 8.7 k$\Omega$ at the charge neutrality point (CNP). Due to the doping not only the CNP shifted to larger values (-42 V for n-type doping and +32 V for p-type doping), but also the resistance at the CNP increased. For better comparability, we define the on/off ratio of a device as a figure of merit by the ratio of the resistance at the CNP and the resistance at a carrier concentration of $10^{13}$ cm$^{-2}$. While the on/off ratio of the undoped device is around 8, the on/off ratios of the n-type and p-type device reach values of 29 and 20, respectively. Previous investigations on double-gated bilayer graphene FETs \cite{Xia,Szafranek} showed similar behavior, which was related to the opening of an energy gap due to symmetry breaking by the perpendicularly applied electric field. 

\begin{figure}[!t]
\includegraphics[width=\linewidth]{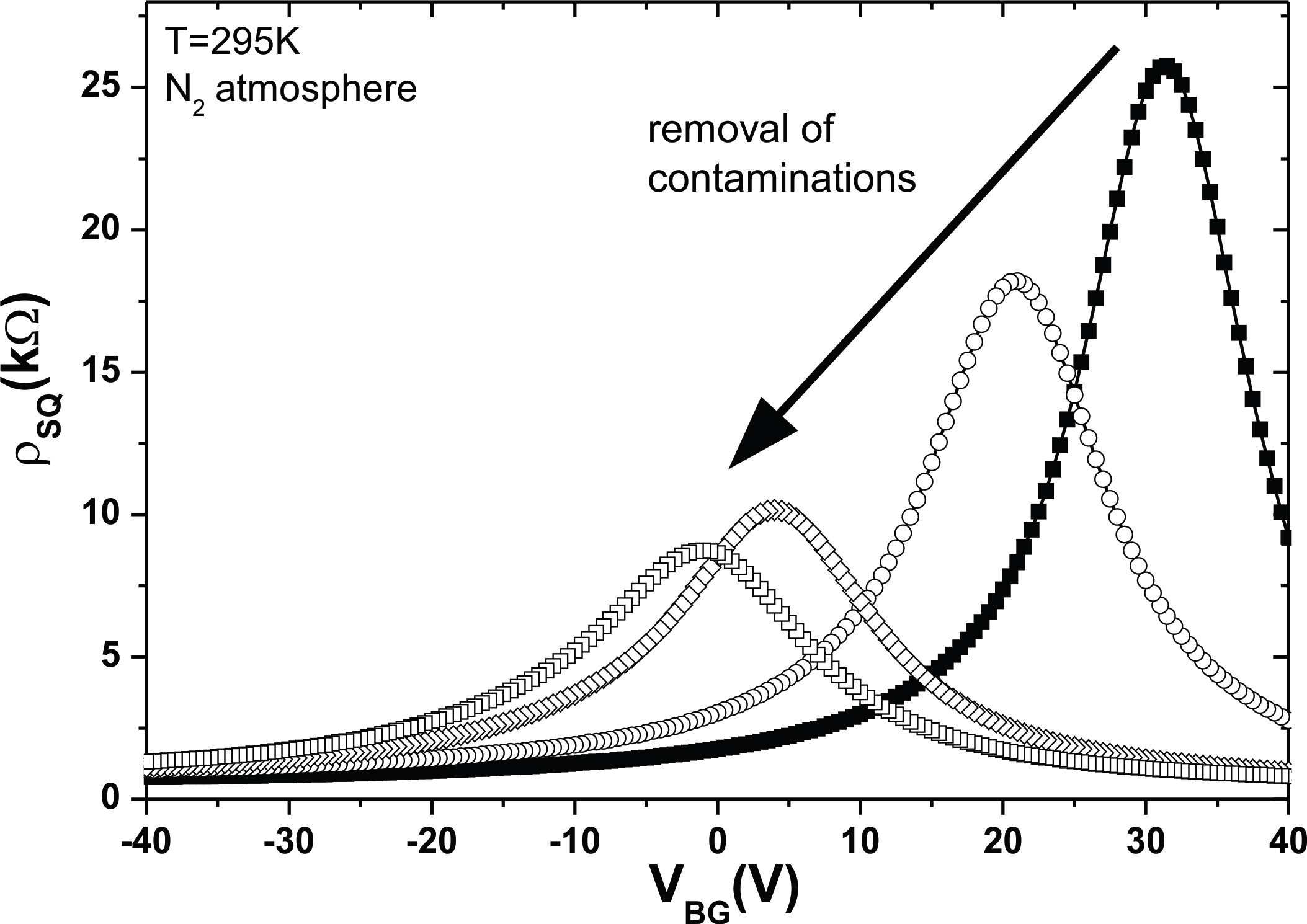}
\caption{Transfer characteristics of a bilayer graphene device after fabrication and atmospheric doping (filled squares) and subsequent annealing at 50~°C (circles), 100~°C (diamonds) and 150~°C (open squares) in nitrogen atmosphere. Measurements were performed at room temperature in nitrogen atmosphere.}
\end{figure}

To ensure that the observed effect is related to the surface dopants on the graphene we annealed several atmospherically doped samples in a nitrogen atmosphere to reduce the doping concentration. Annealing is a reliable method to remove volatile adsorbates from the graphene surface \cite{Lohmann}. Fig.~2 shows the transfer characteristics of a graphene device, where the surface doping was removed by consecutive in-situ annealing steps at 50 °C, 100 °C and 150 °C in nitrogen atmosphere. Due to the stepwise annealing not only the CNP shifted towards zero back gate voltage, but also the resistance at the CNP and the on/off ratio decreased from 26 k$\Omega$ to 8.7 k$\Omega$ and from 20 to 8, respectively. We note that subsequent exposure to ambient atmosphere after annealing, although restoring the initial p-type doping, leads to a significantly lower on/off ratio. The field effect mobility $\mu = \textrm{d}(1/\rho_{SQ})/\textrm{d}V_{BG}\times C_{SQ}^{-1}$ of the sample measured at a carrier concentration of $10^{13}$ cm$^{-2}$ did not change significantly due to the annealing and was approx. 1400 cm$^2$/Vs before and after annealing. Here $C_{SQ}$ denotes the back gate capacitance per area (38 nF/cm$^2$ in our devices). We investigated in total 10 individual samples with atmospheric doping, which all showed similar behavior. In Fig.~3 the on/off ratios of all investigated devices are plotted as a function of their respective back gate voltage at the CNP. For better comparability, we calculate the applied electric displacement field using $D = \epsilon_r V_{BG}/d_{OX}$ with $\epsilon_r$ = 3.9 and oxide thickness d$_{OX}$ = 90~nm. 

The devices doped by a 1 nm thin Al layer all showed a strong n-type doping with the CNP located at back-gate voltages ranging from -20 to -50~V (see Fig. 3). This corresponds to a displacement field $D$ of -0.9 to -2.2~V/nm. In these devices the on/off ratio reached values from 18 to 44, which is slightly larger than the values achieved by atmospheric doping at similar displacement fields. This difference can hardly be explained by a difference in the mobility \cite{Miyazaki}, as the field effect mobility of the Al and atmospheric doped samples was in the same range, both ranging from 900 to 1400 cm$^2$/Vs. Thus we assume that the higher on/off ratio in the aluminium doped devices is related to smaller potential fluctuations and disorder, as these are also expected to be a limiting factor for high on/off ratios in double gated bilayer FETs \cite{Oostinga,Zou,Taychatanapat}. The on/off ratios at a displacement field of 1.7 V/nm obtained by adsorbate doping in the present work ($\approx$40 by aluminum doping) are comparable to those achieved in double gated devices (74 \cite{Xia} and 33 \cite{Szafranek}). Therefore adsorbate doping is an applicable solution for replacing a constantly biased gate electrode in double gated bilayer FETs.

\begin{figure}[!t]
\includegraphics[width=\linewidth]{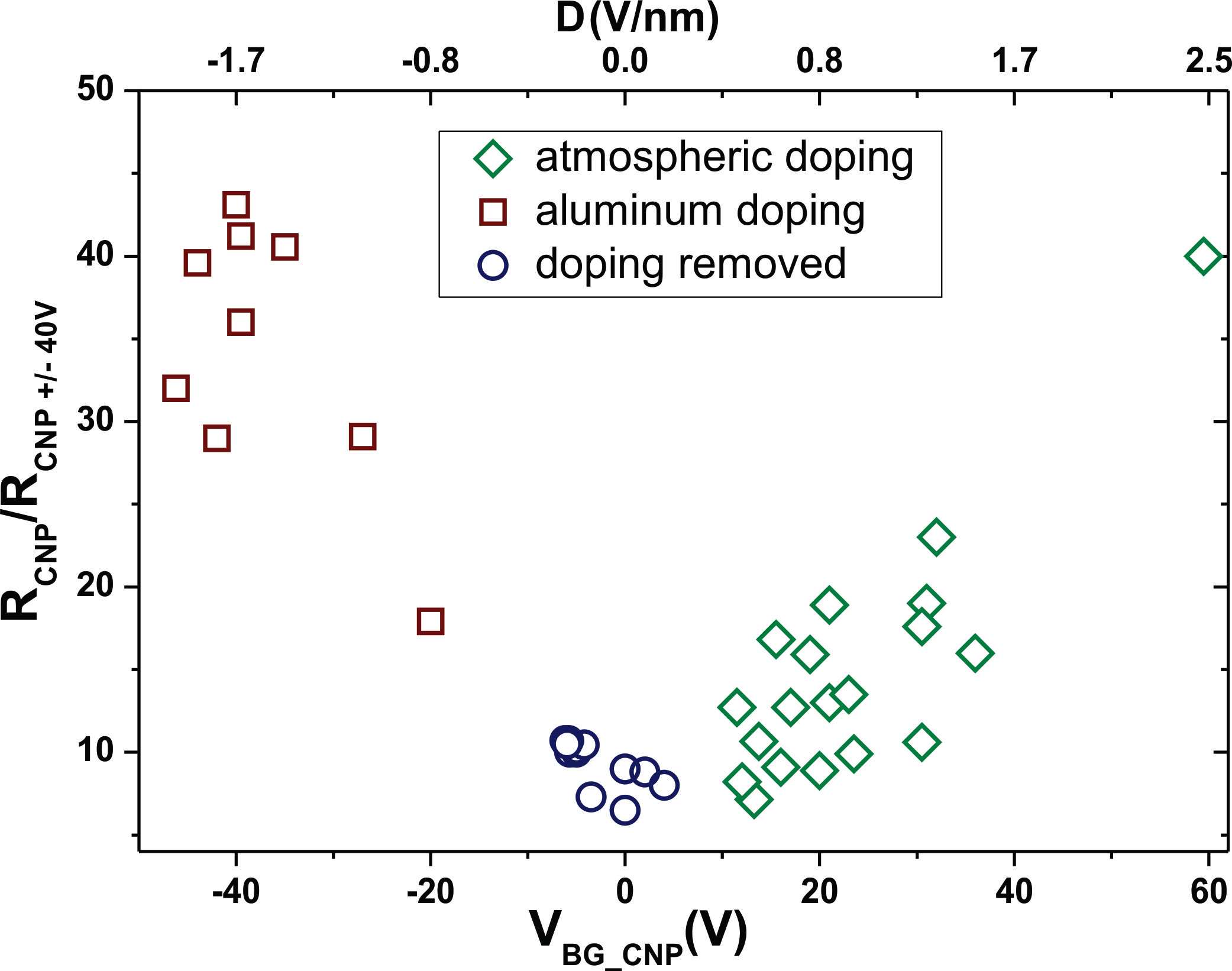}
\caption{On/off-ratio as a function of the back-gate voltage at the charge neutrality point V$_{BG\_CNP}$ of several devices. All measurements have been performed in nitrogen atmosphere and at room temperature.}
\end{figure}

Low temperature transport measurements were performed on three atmospherically doped devices to investigate the physical origin of the increased on/off ratios. Fig.~4 shows the transfer characteristics of a bilayer graphene device at temperatures between 190 and 10~K. At the CNP, we observe an increase of the resistivity with decreasing temperature, while at higher charge carrier densities, i.e. for V$_{BG}$ = 30~V and 0~V, the resistivity is nearly constant. The temperature dependency of the resistance indicates that an energy gap is present at the CNP and that at higher carrier concentration the mobility is independent on the temperature. In the inset of Fig.~4 the normalized resistance at the CNP is shown for three devices as a function of the reciprocal temperature. Starting at room temperature R$_{CNP}$ increases exponentially with decreasing temperature down to around 170~K. This exponential increase is indicated by the sketched straight lines. For temperatures below 170~K the increase of the resistance with decreasing temperatures is weaker than for higher temperatures and starts to saturate. A similar temperature dependence was previously observed in several experiments on double gated bilayer graphene devices (e.g. \cite{Oostinga,Xia,Taychatanapat,Miyazaki,Zou}). Zou \textit{et al.} \cite{Zou} and Miyazaki \textit{et al.} \cite{Miyazaki} argue that at temperatures close to room temperature transport in double gated bilayer graphene is mediated by thermal activation of carriers to the edge of a mobility gap, generated by the applied electric displacement field. Therefore we attribute the high on/off ratios observed in our experiments to the opening of a mobility gap. This mobility gap is caused by a symmetry breaking electric displacement field generated by the adsorbate doping on top and the applied back-gate voltage. 

\begin{figure}[!t]
\includegraphics[width=\linewidth]{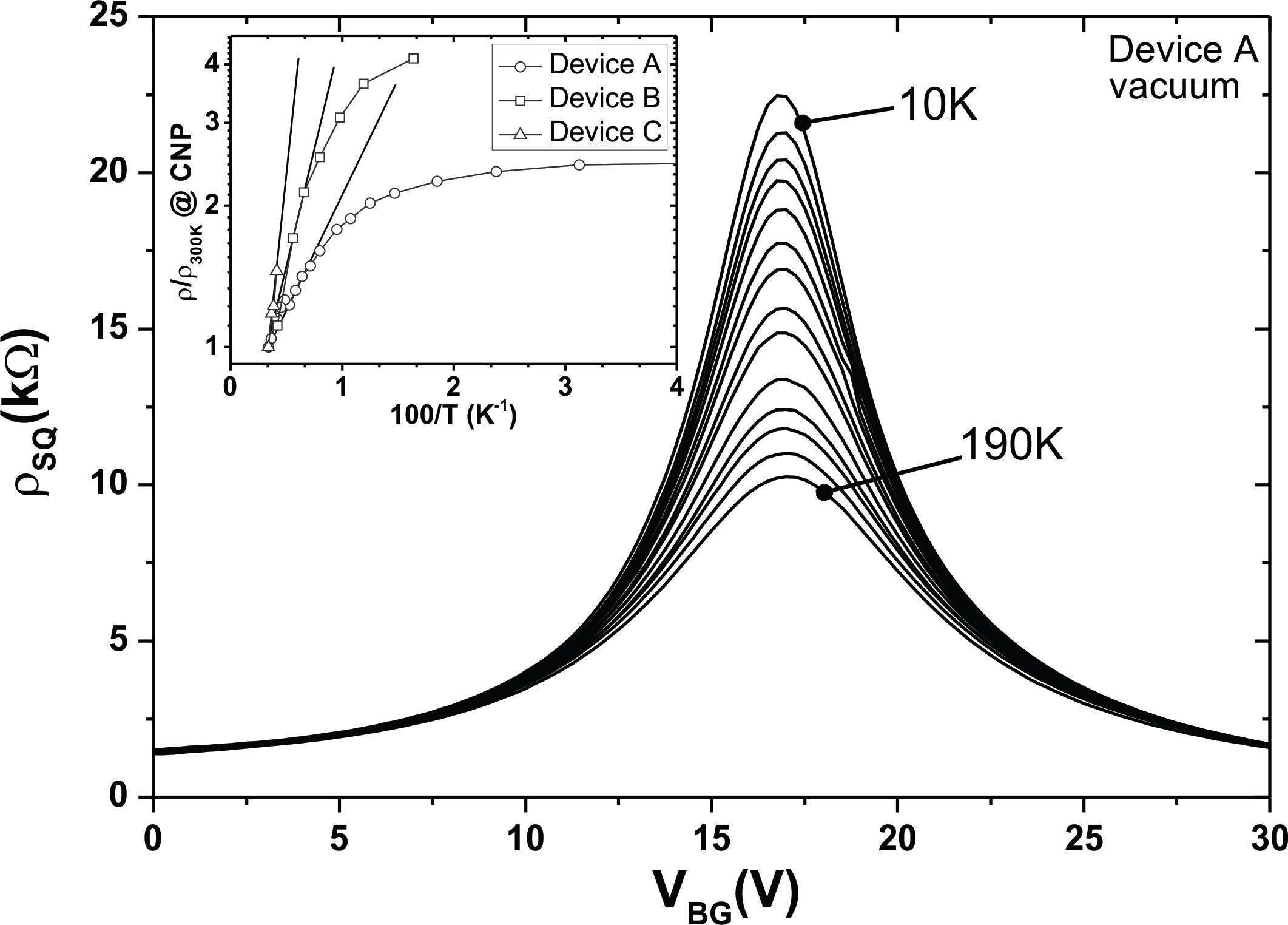}
\caption{Transfer characteristics at temperatures from 190~K down to 10~K. Inset: Normalized resistance at CNP as a function of inverse temperature T$^{-1}$ for three different devices.}
\end{figure}

From the resistance increase at temperatures between 300 and 150~K the size of the mobility gap can be determined. With the sketched straight lines the mobility gap is calculated to values of 40~meV for device A (CNP located at 17 V back gate voltage) and 80~meV for device C (CNP located at 35 V back gate voltage). These values of the mobility gap size are in good accordance with the results obtained on double gated bilayer graphene devices at comparable displacement fields \cite{Xia,Szafranek,Zou}.
Device B shows a mobility gap of 52~meV, which is smaller than expected for a CNP at 57~V ($D$=2.4 V/nm). We attribute this to a degradation of the device through repeated annealing and exposing to ambient air before the low temperature measurements were performed.

In summary, we have investigated the influence of atmospheric and aluminum adsorbate doping on the transport properties in bilayer graphene FETs. We have shown that the on/off ration of bilayer graphene FETs can be increased significantly by adsorbate doping, reaching values comparable to those measured in double gated bilayer FETs. The increased on/off ratio can be attributed to the opening of a mobility gap due to symmetry breaking, similar to double gated bilayer FETs. Therefore adsorbate doping is ideally suited for operating bilayer FETs with only one gate electrode or for realizing more complex devices like bilayer tunneling FETs \cite{Fiori}. Furthermore, our approach paves the way for optoelectronic devices such as THz detectors or ultra sensitive sensors and enables the creation of superlattice structures on graphene by forming an artificial potential landscape.

This work was supported by the European Union under contract number 215752
("GRAND") and by the German Federal Ministry of Education and
Research (BMBF) under contract number NKNF 03X5508 ("ALEGRA").

\end{document}